\documentclass[twocolumn,amsmath,amssymb]{revtex4}
\usepackage{graphicx}
\usepackage{dcolumn}
\usepackage{bm}
\begin{document}

\title{Interaction-Driven Distinctive Electronic States of Artificial Atoms at the ZnO Interface}
\author{Tapash Chakraborty $^1$,\footnote{Tapash.Chakraborty@umanitoba.ca}
Aram Manaselyan$^2$, and Manuk Barseghyan$^2$}
\affiliation{$^1$ Department of Physics and Astronomy, University of
Manitoba, Winnipeg, Canada R3T 2N2}
\affiliation{$^2$ Department of Solid State Physics, Yerevan State
University, Yerevan, Armenia}
\date{\today}
\begin{abstract}
We have investigated the electronic states of planar quantum dots at the ZnO interface containing
a few interacting electrons in an externally applied magnetic field. In these systems, the 
electron-electron interaction effects are expected to be much stronger than in traditional semiconductor 
quantum systems, such as in GaAs or InAs quantum dots. In order to highlight that stronger Coulomb 
effects in the ZnO quantum dots, we have compared the energy spectra and the magnetization in
this system to those of the InAs quantum dots. We have found that in the ZnO quantum dots,
the signatures of stronger Coulomb interaction manifests in an unique ground state that has very different 
properties than the corresponding ones in the InAs dot. Our results for the magnetization also exhibits 
behaviors never before observed in a quantum dot: We have found a stronger temperature dependence and
other unexpected features, such as {\it paramagnetic-like} behavior at high temperatures for a quantum-dot 
helium.
\end{abstract}

\maketitle

For decades, creation of high-mobility two-dimensional electron gas (2DEG) in quantum confined 
semiconductor heterojunctions has paved the way for present-day electronics and quantum devices 
and was crucial for many seminal discoveries in correlated electron systems in a magnetic field. 
The most notable example was the fractional quantum Hall effect \cite{fqhe} and several other 
unique phenomena in various nanoscale systems, viz. the quantum dots (QDs) (or, the {\it artificial 
atoms}) \cite{Qdots,QD_InAs,heitmann} and quantum rings (QRs) \cite{Qrings,haug}. Similar phenomena have 
also been investigated in Dirac materials such as graphene \cite{abergeletal,graphene_book} 
and other graphene-like materials, such as silicene \cite{silicene}, germanene \cite{germanene,buckled} 
and black phosphorous \cite{dressel,phospho}. In recent years, very exciting developments have taken 
place with the creation of high-mobility 2DEG in heterostructures with insulating complex oxides. Unlike 
in traditional semiconductors, electrons in  these systems are strongly correlated \cite{mannhart}. These 
should then exhibit effects ranging from strong electron correlations, magnetism, interface superconductivity, 
tunable metal-insulator transitions, among others, and of course, the exciting possibility of all-oxide
electronic devices. Interestingly, odd-denominator fractional quantum Hall states were discovered in 
MgZnO/ZnO heterojunction \cite{zno_fqhe}, and surprising results were found for the even-denominator states 
\cite{falson,luo}, and in a tilted magnetic field \cite{tilted}. Preparation of various nanostructures, such 
as nanorings, nanobelts, etc. have been reported in ZnO \cite{zno_nano}. Given the enormous potential for this
newly developed source of 2DEG, it is therefore imperative that the electronic properties of quantum confined 
systems at the oxide interfaces are understood.

Here we report on our studies of the electronic states of artificial atoms in this new planar electron system 
with the hope that the strong correlation effects that are expected at the ZnO interface will manifestly alter 
the electronic states in these structures. While the Coulomb interaction strength is strong in this system, the 
Rashba spin-orbit interaction strength \cite{bychkov} is found to be very small \cite{kozuka} for the 2DEG at the ZnO 
interface and therefore will not be important for our present purpose. We have explored the energy spectra of two and 
three electrons in a parabolic confinement and compared these results with those for `conventional' QD systems, 
such as in InAs \cite{QD_InAs}. These two- and three-electron QDs have been thoroughly explored in traditional 
semiconductors by various experimental groups \cite{Qdots,heitmann}. Magnetization of the artificial atoms 
\cite{Qdots,heitmann,QD_mag,Siranush}, (and of quantum rings \cite{Qrings}) is an important probe that reflects entirely 
on the properties of the energy spectra. It is a thermodynamical quantity of the QDs that has received some experimental 
attention \cite{wilde_10,schwarz,delft_98}, particularly after the theoretical prediction that the electron-electron 
interaction is directly responsible for the magnetic field dependence of this quantity \cite{QD_mag}. 
Our study indicates that, stronger electron-electron interaction exerts a very profound influence 
on the electronic states and on magnetization of the ZnO QD. As an example, in a three-electron QD, even 
in the absence of the magnetic field, the ground state is changed to $L=0$, $S=-3/2$, instead of the
expected state at $\vert L\vert=1$, and $S=\frac12$, that is found, for example, in the case of the InAs QD. 
This interesting and unexpected result would manifests itself in optical and magnetic characteristics of the 
ZnO QDs. Our results for the magnetization of the ZnO QDs display the expected step-like behavior. But in 
contrast to those of the InAS QDs, the corresponding jumps in the magnetization are observed for much smaller values 
of the magnetic field. The magnetization of the ZnO QDs reveals a very strong temperature dependence, and for 
a two-electron QD, the magnetization reveals a monotonic increase with the magnetic field, much akin to
a paramagnetic system.

Our study here involves a two dimensional QD with cylindrical symmetry, based on the 2DEG
at the ZnO interface, containing few electrons, in a magnetic field that is applied in the growth
direction. The Hamiltonian of our system is
\begin{equation}
{\cal H}=\sum_i^{N^{}_e}{\cal H}_\mathrm{SP}^i+\frac12\sum_{i\neq j}^{N^{}_e}V^{}_{ij}, \label{Ham2D}
\end{equation}
where $N^{}_e$ is the number of electrons in QD, $V^{}_{ij}=e^2/\epsilon\left|\mathbf{r}^{}_i
-\mathbf{r}^{}_j\right|$ is the Coulomb interaction term, with dielectric constant of the dot
material $\epsilon$, and ${\cal H}^{}_\mathrm{SP}$ is the single-particle Hamiltonian in the 
presence of an external perpendicular magnetic field.
\begin{equation}\label{Hsp}
{\cal
H}^{}_\mathrm{SP}=\frac{1}{2m}\left(\textbf{p}-\frac ec\textbf{A}\right)^2+\frac12m
\omega_0^2r^2+\frac12 g\mu^{}_BB\sigma^{}_z,
\end{equation}
where $\textbf{A}=B/2(-y,x,0)$ is the vector potential, and $m$ is the electron effective
mass. We choose the confinement potential of QD as parabolic with parameter $\omega^{}_0=\hbar/{mR^2}$, 
where $R$ is the radius of the dot. The last term of (\ref{Hsp}) is the Zeeman splitting. The 
eigenfunctions of the single-electron hamiltonian (\ref{Hsp}) are the Fock-Darwin
orbitals \cite{Qdots} $f^{}_{nl}(r,\theta)$, where $n$, $l$ are the radial and angular 
quantum numbers. In order to evaluate the energy spectrum of the many-electron system, we need to
digonalize the matrix of the Hamiltonian (\ref{Ham2D}) in a basis of the slater determinants 
constructed from the single-electron wave functions \cite{Qdots,AramRaman}. 
Our investigations were carried out for the ZnO QD with parameters $m=0.24m^{}_0$, $g=4.3$,
$\epsilon=8.5$ \cite{Handbook}. For comparison we have also presented the corresponding
results for the InAs QD with parameters $m=0.042m^{}_0$, $g=-14$, $\epsilon=14.6$ respectively
\cite{QD_InAs,Siranush}. Here we have considered the QDs to be of same sizes with radius $R$.
This corresponds to different values of the confinement potential parameter $\hbar\omega^{}_0$ in the
two cases. For the ZnO QD $\hbar\omega^{}_0=1.5$meV while for the InAs QD $\hbar\omega^{}_0=7.5$meV.

\begin{figure}
\includegraphics[width=8.5cm]{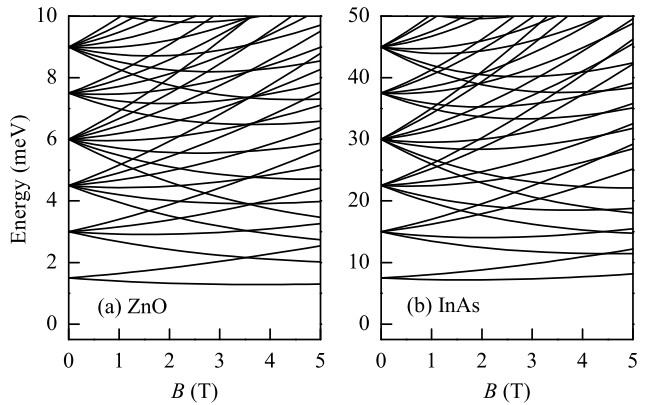}
\caption{\label{fig:EdepB1} The Fock-Darwin energy spectra for (a) ZnO quantum dot, and (b) InAs quantum dot.}
\end{figure}

In Fig.~\ref{fig:EdepB1} the Fock-Darwin spectra for the ZnO (a) and InAs (b) QDs are presented. It is clear 
from the figure that without the magnetic field for the ZnO QD, the energy values are lower and 
the levels are closer to each other due to the larger value of electron effective mass. For the ZnO QD the 
ground state is $l=0$, $s=-1/2$ for all the values of magnetic field. In contrast to that, the ground state 
for the InAs QD has $l=0$, $s=1/2$. For the excited states many level crossings are visible. For the 
ZnO QD these crossings are observed for lower values of magnetic field.

At zero temperature the magnetization $\cal M$ of the QD is defined as ${\cal M}=-\frac{\partial 
E^{}_0}{\partial B}$, where $E^{}_0$ is the ground state energy of the many-electron system 
\cite{Siranush}. Here we report on our studies of the magnetic field dependence of $\cal M$ by evaluating 
the expectation values of the magnetization operator $\widehat{m}=-\frac{\partial {\cal H}}{\partial 
B}$, where ${\cal H}$ is the system Hamiltonian (\ref{Ham2D}). We then need to evaluate the expectation 
values of magnetization operator $\widehat{m}$ using the wave functions of the interacting 
many-electron system. We have also studied the temperature effect on magnetization, following the 
thermodynamical model discussed in \cite{Sir25}. The temperature dependence of magnetization is 
evaluated from the thermodynamic expression \cite{Sir25}
\begin{equation}
{\cal M}=-\displaystyle{\sum^{}_i\frac{\partial E^{}_i}{\partial
B}e^{-E^{}_i/kT}}/{\displaystyle{\sum^{}_i e^{-E^{}_i/kT}}},
\end{equation}
where the partial derivatives are evaluated as expectation values of operator $\widehat{m}$ for the 
interacting state $i$.

\begin{figure}
\includegraphics[width=8.5cm]{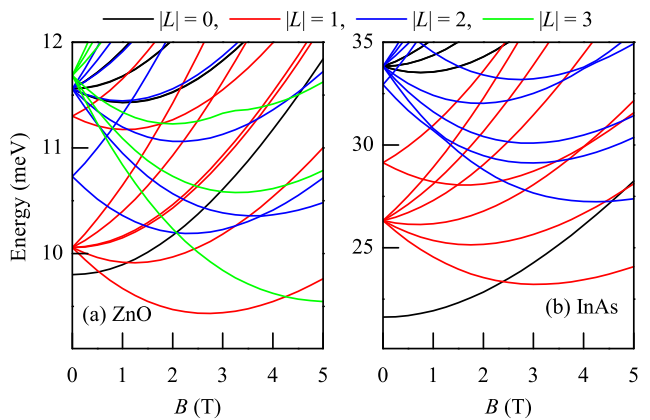}
\caption{\label{fig:EdepB2} The dependences of the low-lying energy levels on the magnetic field 
$B$ for two interacting electrons in the (a) ZnO quantum dot, and (b) InAs quantum dot.}
\end{figure}

In Fig.~\ref{fig:EdepB2}, several low-lying energy levels for the ZnO and InAs QDs with two electrons are presented 
against the magnetic field for various values of total angular momentum $L$. These figures clearly
indicate that in the absence of the magnetic field for the ZnO QD, the energy values are lower 
and the levels are closer to each other due to the larger value of the electron effective mass in the
former case. It is well known that for small values of the magnetic field the ground state of a 
two-electron QD is a singlet state with total angular momentum $L=0$ and spin $S=0$. With an increase 
of the magnetic field a singlet-triplet transition of the ground state is observed. For the 
InAs QD this transition occurs at $B=2.4$T and the ground state changes to triplet state with $|L|=1$, 
$S=1$. For the ZnO QD we can observe a similar transition but for a much smaller value of the 
magnetic field, i.e., at $B=0.55$T which will change the ground state to the triplet state with $|L|=1$, 
$S=-1$.  The spin difference of the two triplet ground states can be explained by the sign of the
g-factor in the two cases. Due to the strong Coulomb interaction, with further increase of the magnetic 
field, a {\it second ground state transition} is observed for the ZnO QD at $B=4.2$T which changes 
the ground state to $|L|=3$, $S=-1$. This transition dose not occur in the case of the InAs QD for 
experimentally observable ranges of the magnetic field.

\begin{figure}
\includegraphics[width=8.5cm]{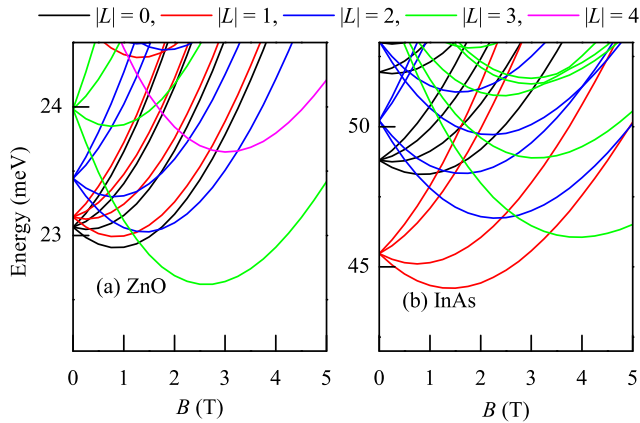}
\caption{\label{fig:EdepB3} The dependence of the low-lying energy levels on the magnetic field $B$ 
for three electrons in (a) ZnO quantum dot, and (b) InAs quantum dot.}\end{figure}

In Fig.~\ref{fig:EdepB3}, the dependencies of low-lying energy levels of the ZnO and InAs QDs with three interacting
electrons are presented against the magnetic field for various values of the total angular momentum $L$. 
Usually in QDs for small values of the magnetic field the three electron ground state has total angular
momentum $|L|=1$ and total spin $S=1/2$ which was observed by many authors in the case of InAs, GaAs and other 
QDs \cite{Qdots,heitmann}. But our present study indicates that due to the strong Coulomb correlation effects in 
ZnO QD, for small values of the magnetic field the three-electron ground state has the total angular momentum 
$L=0$ and total spin $S=-3/2$. With the increase of the magnetic field, at $B=1.3$T again a ground state 
transition can be observed to the state with $|L|=3$ and $S=-3/2$. Similar ground state transition is 
also observed for InAs QD with three electrons, but for a larger value of the magnetic field, viz.,
at $B=3.4$T. These interesting results will manifest themselves in optical and magnetic characteristics 
of the ZnO QDs. In particular, we have considered here the magnetization of the ZnO QD with few electrons 
and compared our results with similar ones for the InAs QD.

\begin{figure}
\includegraphics[width=8.5cm]{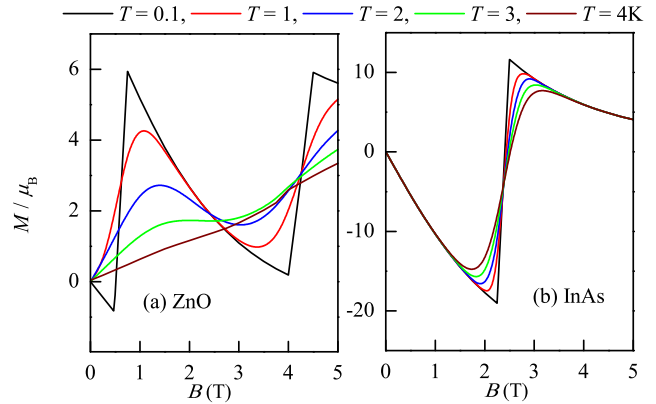}
\caption{\label{fig:Mag2} The magnetization of two electron quantum dot for various values of the
temperature for (a) ZnO QD and (b) InAs QD.}
\end{figure}

\begin{figure}
\includegraphics[width=8.5cm]{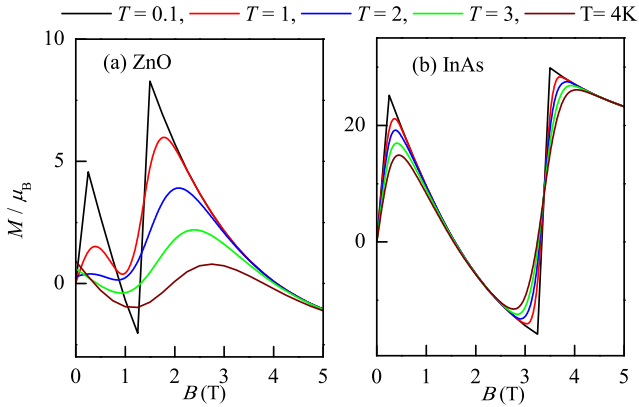}
\caption{\label{fig:EdepBMub8nPhi} The magnetization of the three-electron quantum dot for various values of 
the temperature for (a) ZnO QD and (b) InAs QD.}
\end{figure}

In Fig.~\ref{fig:Mag2} the magnetization of the ZnO and InAs QDs with two electrons are presented against the applied
magnetic field for various values of the temperature from 0.1 to 4K. Similar results but for three electron 
ZnO and InAs QDs are presented in Fig.~\ref{fig:EdepBMub8nPhi}. In all cases, the magnetic field dependencies of 
the magnetization at $T=0.1$K have step like behaviors. These jumps in the magnetization can be explained by the ground 
state oscillations of few electron QDs \cite{QD_mag}. In the case of the two-electron ZnO QD (Fig.~\ref{fig:Mag2}(a)) there
are two jumps in the magnetization for magnetic fields $B=0.55$T and 4.2T. This result is to be contrasted
to the case of the two-electron InAs QD, where only one jump is observed at $B=2.4$T. Also it should be noted 
that the magnetization of ZnO QD has a very strong temperature dependence as compared to that of the InAs
quantum dots. Due to the large value of the electron effective mass in ZnO the excited energy levels of 
the few electron QD are very close to ground state and the increase of temperature mixes these states 
that explains the  smoothening and averaging of the magnetization curves. Furthermore, for 
$T=4$K a surprising {\it paramagnetic-like} behavior of the magnetization is observed for the two-electron 
ZnO QD. In Contrast to the ZnO QD, the magnetization curves of the InAs QDs have a weak temperature
dependence (Fig.~\ref{fig:Mag2}(b)). Similar behaviors are also observed for three-electron ZnO and InAs QDs 
(Fig.~\ref{fig:EdepBMub8nPhi}(a) and Fig.~\ref{fig:EdepBMub8nPhi}(b)). Here again a step like behavior of the 
magnetization is observed for both QDs. At zero temperature the first jump is caused by the lifting of fourfold 
degeneracy of the ground state due to the magnetic field. The second jump for the ZnO QD at $B=1.4$T is caused by 
the change of the ground state.

\begin{figure}
\includegraphics[width=8.5cm]{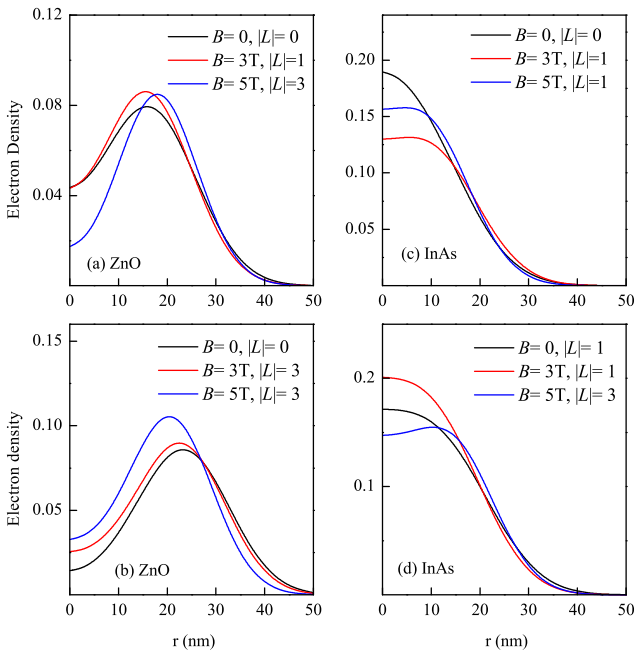}
\caption{\label{fig:density} Ground state density for various values of magnetic field for (a) two-electron
ZnO QD, (b) three-electron ZnO QD, (c) two-electron InAs QD, and (d) three-electron InAs QD.
}
\end{figure}

In Fig.~\ref{fig:density} the ground state density is presented for two- and three-electron ZnO QDs 
and InAs QDs for various values of the magnetic field. In a InAs QD, the electrons are mostly 
located in the central part of the dot, while for the ZnO QD, due to weaker confinement and stronger Coulomb 
interaction, electrons are repelled from the central part of the dot. With an increase of the magnetic field 
the ground state electron densities for the ZnO QD and the InAs QD exhibit completely different behaviors which 
can be explained by the different ground state changes discussed above. For example, at $B=0$ the two-electron 
ground state for both QDs is with $L=0$, but at $B=5$T the ground state for the ZnO QD has $|L|=3$ and for the 
InAs QD $|L|=1$. For three electron QDs the ground states are different even at zero magnetic field.

To summarize, we have presented here a detailed and accurate studies of the electronic states and magnetization 
of the ZnO quantum dot with few interacting electrons in an externally applied magnetic field. Our results have 
been compared with results for the InAs QD to highlight the unique features of the ZnO quantum dot.  
We have shown that electron-electron interaction exerts a very strong influence on the electronic 
states and on magnetization of the ZnO QD. In particular, the energy levels of a two-electron ZnO QD display more
level crossing for finite values of the magnetic field. Additionally, in the case of the three-electron QD, even in
the absence of the magnetic field the ground state is changed to $L=0$, $S=-3/2$, as compared to the usual
state at $\vert L\vert=1$, and $S=\frac12$, that is found, for example, for the InAs QD. These interesting 
and unexpected results will manifest itself in optical and magnetic characteristics of the ZnO QDs. Further, 
we have shown that the magnetization curves of the ZnO QDs have the expected step-like behavior, but in contrast 
to the InAS QDs, the corresponding jumps in magnetization are observed for much smaller values of the magnetic field. 
Therefore the ZnO QDs are suitable for low-field magnetization measurements. The magnetization of the ZnO QDs has 
a very strong temperature dependence, and surprisingly, at high temperatures, the two-electron ZnO QD shows a 
paramagnetic-like behavior.

The work has been supported by the Canada Research Chairs Program of the Government of Canada, the Armenian State 
Committee of Science (Project no. 15T-1C331) and Armenian National Science and Education Fund (ANSEF Grant no. nano-4199).

\end{document}